\documentstyle[12pt,psfig,epsf]{article}
\begin{document}
\null
\vspace{3cm}

\begin{center}
{\large\bf RELATIVISTIC EFFECTS IN THE ELECTROMAGNETIC
STRUCTURE OF  $\rho$-MESON}\\

\vspace{4mm}
A.F.\,Krutov\footnote[1]{Krutov Alexander Fedorovich, Dept. of General and
Theoretical Physics, Samara State University, e--mail: krutov@ssu.samara.ru}
V.E.\,Troitsky\footnote[2]{Troitsky Vadim Eugenievich, Dept. of Theoretical
High Energy Physics, D.V.Skobeltsyn Institute of Nuclear Physics, Moscow
State University,
e--mail: troitsky@theory.sinp.msu.ru}
\end{center}

\begin{abstract}
The relativistic calculations of the electromagnetic form factors and
static moments of $\rho$-meson are given in the framework of the
relativistic Hamiltonian dynamics with  different model wave functions.
The impulse approximation is used. Lorentz covariance and conservation law
for the electromagnetic current operator are taken into account.

\end{abstract}
\section*{Introduction}

A new relativistic approach to electroweak properties of
composite systems has been proposed in our recent papers
\cite{KrT02,KrT03prc}. The approach is based on the use of the instant
form (IF) of relativistic Hamiltonian dynamics (RHD).
The detailed description of RHD can be found in the review
\cite{KeP91}.
Some other
references as well as some basic equations of RHD
approach are given in  Ref.\cite{KrT02,KrT03prc}.

In the paper  \cite{KrT02} our approach was used to give a
realistic calculation of electroweak properties of pion considered as
composite quark--antiquark system. The electromagnetic form factor and the
lepton decay constant were calculated for pion using different
model wave functions for the relative motion of quarks in pion.

Now our aim is to describe the electromagnetic properties of
more complicated systems, namely, $\rho$-meson as
the composite system of two particles of
spin 1/2 (quarks) with total spin one and
total angular momentum one  and with zero orbital momentum.
The main problem is
a construction of electromagnetic current operator satisfying
standard conditions (Lorentz covariance, conservation law
etc., see, e.g., Ref.\cite{Lev95}).

The basic point of our approach
\cite{KrT02,KrT03prc} to the construction of the electromagnetic current
operator is the general method of relativistic invariant parameterization of
local operator matrix elements proposed as long ago as in
1963 by Cheshkov and Shirokov~\cite{ChS63}.

In fact, this parameterization is a
realization of the Wigner--Eckart theorem for the Poincar\'e
group and so it enables one for given matrix element of arbitrary
tensor dimension to separate the reduced
matrix elements (form factors) that are invariant
under the Poincar\'e group. The matrix element of a
given operator is represented as a sum of terms, each one of
them being a covariant part multiplied by an invariant part.
In such a representation a covariant part describes
transformation (geometrical) properties of the matrix element,
while all the dynamical information on the transition is
contained in the invariant part -- reduced matrix elements.
In the case of composite systems these form
factors appearing through the canonical parameterization are to
be considered in the sense of distributions, that is they are
generalized instead of classical functions. As was demonstrated
in Ref.\cite{KrT02} this fact takes place even in nonrelativistic
case. It is in terms of form factors that the electroweak
properties of composite systems are described in the frame of
the approach \cite{KrT02,KrT03prc}.

In our approach some rather general problems arising in the
description of composite quark models have been solved. For example,
the description of electromagnetic properties of composite
systems in terms of form factors in Ref.\cite{KrT02}, in fact,
solves the problem of construction of the electromagnetic
current satisfying the conditions of translation invariance,
Lorentz covariance, conservation law, cluster separability and
equality of the composite system charge to the sum of constituents charges
(charge nonrenormalizability).

Let us note that the importance of the problem of
construction of electromagnetic current is
actual not only for RHD but for all relativistic approaches to
composite systems, including the field theoretical approaches
\cite{Lev95,GrR87,ChC88,VaD95,LeP98,MeS98,Kli98}.

Our calculations of the electromagnetic characteristics of $\rho$-meson
are performed in
the well known relativistic impulse approximation (IA).
It means that the electromagnetic current of a composite system
is a sum of one--particle currents of the constituents. It is
worth emphasizing that in our method this
approximation does not violate the standard conditions
for the current. To--day a construction of
relativistic impulse approximation without breaking of
relativistic covariance and current conservation law is a common
trend of different approaches
\cite{Lev95,VaD95,LeP98,Kli98}.
Let us note that in the present paper it is for the first time
that such a construction has been realized for the case of nonzero
spin in the frame of IF RHD. This is a variant of the
relativistic impulse approximation formulated in terms of
reduced matrix elements (see Ref.~\cite{KrT02}) -- the modified
impulse approximation (MIA).

Using different model wave functions
of quark relative motion we calculate the electromagnetic
form factors and the static properties of $\rho$-meson
supposing quarks to be in the $S$ state of relative motion. It
is interesting to mention that relativistic effects occur to
produce a nonzero quadrupole moment and quadrupole form
factor. It is well known that in the nonrelativistic case the
non--zero quadrupole form factor is caused by the presence of
the $D$- wave and is zero otherwise.

The paper is organized as follows.
In Sec.I the formalism developed in papers \cite{KrT02,KrT03prc}  is used
in the case of the system with total spin one and total angular momentum
one. In terms of reduced matrix elements the so called modified impulse
approximation (MIA) ~\cite{KrT02} is formulated. In MIA $\rho$ -- meson
form factors are obtained explicitly.

In Sec.II the results of calculations of static properties
and electromagnetic form factors of
$\rho$ meson are discussed.

In Sec.III the conclusion is given.

\section{Integral representation for the
$\rho$-meson electromagnetic form factors}

\renewcommand{\theequation}{\arabic{equation}}

Let us consider the matrix element of the $\rho$-meson electromagnetic
current.
In our composite quark model the $u$- and $\bar d$-- quarks are
in the $S$ state of relative motion, that is $l = l' =$ 0,
with total spin one and
total angular momentum one: $J=J'=S=S'=$1. This matrix element
is given in the paper \cite{KrT03prc}:
\begin{equation}
\langle\vec p_c\,,m_{Jc}|j_\mu(0)|\vec p_c\,'\,,m'_{Jc}\rangle
= \langle\,m_{Jc}|\,D^{1}(p_c\,,p'_c)\,
\sum_{i=1,3}\,
\tilde{\cal F}\,^i_c(t)\,\tilde A^i_\mu\,|m'_{Jc}\rangle\;;
\label{jc=FA}
\end{equation}
$$
\tilde{\cal F}\,^1_c(t) = \tilde f^c_{10} +
\tilde f^c_{12}\left\{[i{p_c}_\nu\,\Gamma^\nu(p'_c)]^2
\right.
\left. - \frac{1}{3}\,\hbox{Sp}[i{p_c}_\nu\,\Gamma^\nu(p'_c)]^2\right\}
\frac{2}{\hbox{Sp}[{p_c}_\nu\,\Gamma^\nu(p'_c)]^2}\;,
$$
\begin{equation}
\tilde{\cal F}\,^3_c(t) = \tilde f^c_{30}\;;
\label{FicAic}
\end{equation}
$$
\tilde A^1_\mu = (p_c + p'_c)_\mu\;,\quad
\tilde A^3_\mu =
\frac{i}{M_c}
\varepsilon_{\mu\nu\lambda\sigma}
\,p_c^\nu\,p'_c\,^\lambda\,\Gamma^\sigma(p'_c)\;,
$$
here $p'_c\,,p_c$ are 4-vectors of  $\rho$-meson in initial and final
states, $m'_{Jc}\,,m_{Jc}$ are projections of the total angular momenta,
$D^{1}(p_c\,,\,p'_c)$ is matrix of Wigner rotation,
$\Gamma^\nu(p'_c)$ is 4-vector of spin,
$M_c$ is $\rho$-meson mass,
$\tilde f^c_{10}\,,\,\tilde f^c_{12}\,,$
$\tilde f^c_{30}$ are
charge, quadrupole and magnetic form factors of $\rho$-meson
respectively.

The representation (\ref{jc=FA}), and (\ref{FicAic})
of the matrix element satisfies all the conditions for the
composite system electromagnetic current \cite{Lev95}.

The integral representations for the composite system form factors in
Eq.~(\ref{jc=FA}), and (\ref{FicAic}) are obtained in the papers
\cite{KrT02,KrT03prc}:
\begin{equation}
\tilde f^c_{in}(Q^2) =  \int\,d\sqrt{s}\,d\sqrt{s'}\,
\varphi(s)\,\tilde G_{in}(s,Q^2,s')\varphi(s')\;,
\label{intrepJ1}
\end{equation}
here $\varphi(s)$ is wave function of quarks in RHD,
$\tilde G_{in}(s,Q^2,s')$ is Lorentz covariant generalized function
(reduced matrix element on the Poincar\'e group).

Our form factors in Eq.~(\ref{FicAic}) can be written in terms
of standard Sachs form factors for the system with the total
angular momentum one.  To do this let us write the
parameterization of the electromagnetic current matrix element
in the Breit frame (see, e.g., Ref.~\cite{ArC80}):
$$
\langle\vec p_c\,,m_J|j_\mu(0)|\vec p_c\,'\,,m'_J\rangle =
G^\mu(Q^2)\;,
$$
$$
G^0(Q^2) = 2p_{c0}\left\{(\vec\xi\,'\vec\xi\,^*)\,G_C(Q^2) +
\left[(\vec\xi\,^*\vec Q)(\vec\xi\,'\vec Q) - \frac{1}{3}Q^2
(\vec\xi\,'\vec\xi\,^*)\right]\,\frac{G_Q(Q^2)}{2M_c^2}\right\}\;,
$$
\begin{equation}
\vec G(Q^2) = \frac{p_{c0}}{M_c}\left[\vec\xi'\,(\vec\xi\,^*\vec Q) -
\vec\xi\,^*(\vec\xi\,'\vec Q)\right]\,G_M(Q^2)\;.
\label{Gi}
\end{equation}
Here $G_C\;,\;G_Q\;,\;G_M$  are the charge, quadrupole and
magnetic form factors, respectively.

The polarization vector in the Breit frame has the
following form:
\begin{equation}
\xi^\mu(\pm 1) = \frac{1}{\sqrt{2}}(0\;,\;\mp 1\;,\;-\,i\;,\;0)\;,\quad
\xi^\mu(0) = (0\;,\;0\;,\;0\;,\;1)\;.
\end{equation}
The variables in $\xi$ are total angular momentum
projections.

In the Breit frame:
\begin{equation}
q^\mu = (p_c - p'_c)^\mu = (0\;,\;\vec Q)\;,\quad
p_c^\mu = (p_{c0}\;,\;\frac{1}{2}\vec Q)\;,\quad
p'_c\,^\mu = (p_{c0}\;,\;-\frac{1}{2}\vec Q)\;,
\end{equation}
$$
p_{c0} = \sqrt{M_c^2 + \frac{1}{4}Q^2}\;,\quad
\vec Q = (0\;,\;0\;,\;Q)\;.
$$
Comparing Eq.~(\ref{jc=FA}) with Eq.~(\ref{Gi}) and taking into
account the fact that in the Breit system
$D^1_{m_J\,m''_J}(p_c\,,p_c') = \delta_{m_J\,m''_J}\;,$
we have:
$$
G_C(Q^2) = \tilde f^c_{10}(Q^2)\;,\quad
G_Q(Q^2) = \frac{2\,M_c^2}{Q^2}\,\tilde f^c_{12}(Q^2)\;,
$$
\begin{equation}
\quad G_M(Q^2) = -\,M_c\,\tilde f^c_{30}(Q^2)\;.
\label{GF}
\end{equation}

Let us use for (\ref{intrepJ1}) the modified impulse approximation
formulated in terms of form factors
$\tilde G_{iq}(s,Q^2,s')$.
The physical meaning of this approximation is considered in
detail in Ref.~\cite{KrT02}. In the frame of MIA
the invariant form factors $\tilde G_{iq}(s,Q^2,s')$
in (\ref{intrepJ1}) are changed by the free two--particle invariant
form factors $g_{0i}(s,Q^2,s')\;(i=C,Q,M)$
describing the electromagnetic properties of the system of
two free particles. So, the equations to be used for the
calculation of the $\rho$--meson properties in MIA are the following:
$$
G_C(Q^2) =
\int\,d\sqrt{s}\,d\sqrt{s'}\, \varphi(s)\,g_{0C}(s\,,Q^2\,,s')\,
\varphi(s')\;,
$$
\begin{equation}
G_Q(Q^2) =
\frac{2\,M_c^2}{Q^2}\,\int\,d\sqrt{s}\,d\sqrt{s'}\,
\varphi(s)\,g_{0Q}(s\,,Q^2\,,s')\,\varphi(s')\;,
\label{GqGRIP}
\end{equation}
$$
G_M(Q^2) =-\,M_c\,\int\,d\sqrt{s}\,d\sqrt{s'}\,
\varphi(s)\,g_{0M}(s\,,Q^2\,,s')\,
\varphi(s')\;.
$$

The free two--particle invariant form factors can be calculated by the
methods of relativistic kinematics and have the following form.\\

The charge free two--particle form factor:
$$
g_{0C}(s, Q^2, s') = \frac{1}{3}\,R(s, Q^2, s')\,Q^2\,\times
$$
$$
\times\left\{(s + s'+ Q^2)(G^u_E+G^{\bar d} _E)\,
\left[2\,\cos(\omega_1-\omega_2) + \cos(\omega_1+\omega_2)\right] -
\right.
$$
\begin{equation}
- \left.\frac{1}{M}\xi(s,Q^2,s')(G^u_M+G^{\bar d}_M)\,
\left[2\,\sin(\omega_1-\omega_2) - \sin(\omega_1+\omega_2)\right]\right\}\;.
\label{g0C}
\end{equation}

The quadrupole free two--particle form factor:
$$
g_{0Q}(s, Q^2, s') = \frac{1}{2}\,R(s, Q^2, s')\,Q^2\,\times
$$
$$
\times\left\{(s + s'+ Q^2)(G^u_E+G^{\bar d}_E)\,
\left[\cos(\omega_1-\omega_2) - \cos(\omega_1+\omega_2)\right] -
\right.
$$
\begin{equation}
- \left.\frac{1}{M}\xi(s,Q^2,s')(G^u_M+G^{\bar d}_M)\,
\left[\sin(\omega_1-\omega_2) + \sin(\omega_1+\omega_2)\right]\right\}\;.
\label{g0Q}
\end{equation}

The magnetic free two--particle form factor:
$$
g_{0M}(s, Q^2, s') = -\,{2}\,R(s, Q^2, s')\,\times
$$
$$
\times\left\{\xi(s,Q^2,s')(G^u_E+G^{\bar d}_E)\,
\sin(\omega_1-\omega_2) +
\right.
$$
$$
+ \frac{1}{4M}(G^u_M+G^{\bar d}_M)
\left[(s + s' +Q^2)\,Q^2\left(\frac{3}{2}\cos(\omega_1-\omega_2) +
\frac{1}{2}\cos(\omega_1+\omega_2)\right)\right. -
$$
$$
- \frac{1}{4}\xi(s,Q^2,s')\left[
\beta(s,Q^2,s') + \beta(s',Q^2,s)\right]\times
$$
$$
\times\left[\sin(\omega_1-\omega_2) - \sin(\omega_1+\omega_2)\right]
$$
$$
- \frac{1}{2}\xi^2(s,Q^2,s')\left[
\frac{1}{\sqrt{s'}(\sqrt{s'}+2\,M)}  \right. +
$$
\begin{equation}
\left.\left.\left. + \frac{1}{\sqrt{s}(\sqrt{s}+2\,M)}\right]
\left[\cos(\omega_1-\omega_2) + \cos(\omega_1+\omega_2)\right]\right]\right\}
\;.
\label{g0M}
\end{equation}
In Eqs. (\ref{g0C})--(\ref{g0M}):
$$
R(s, Q^2, s') = \frac{(s + s'+ Q^2)}{2\sqrt{(s-4M^2) (s'-4M^2)}}\,
\frac{\vartheta(s,Q^2,s')}{{[\lambda(s,-Q^2,s')]}^{3/2}}
\frac{1}{\sqrt{1+Q^2/4M^2}}\;,
$$
$$
\xi(s,Q^2,s')=\sqrt{ss'Q^2-M^2\lambda(s,-Q^2,s')}\;,
$$
$$
\beta(s,Q^2,s') =
\frac{(\sqrt{s'}+2\,M)(s-s'+Q^2)+(s'-s+Q^2)\sqrt{s'}}
{\sqrt{s'}(\sqrt{s'}+2\,M)}\;,
$$
$\omega_1$ and $\omega_2$ are the Wigner rotation parameters:
$$
\omega_1 =
\hbox{arctg}\frac{\xi(s,Q^2,s')}{M\left[(\sqrt{s}+\sqrt{s'})^2 +
Q^2\right] + \sqrt{ss'}(\sqrt{s} +\sqrt{s'})}\;,
$$
\begin{equation}
\omega_2 = \hbox{arctg}\frac{
\alpha (s,s') \xi(s,Q^2,s')} {M(s + s' + Q^2)
\alpha (s,s')
+ \sqrt{ss'}(4M^2 + Q^2)}\;,
\label{omega}
\end{equation}
$\alpha (s,s') = 2M + \sqrt{s} + \sqrt{s'} $,
$\vartheta(s,Q^2,s')=
\theta(s'-s_1)-\theta(s'-s_2)$, $\theta$ is the step--function:
$$
\theta(x) = \left\{
\begin{array}{lll}
0\;,&x\;\le\;0\\
1\;,&x\;>\;0\;,\\
\end{array}
\right.
$$
$$
s_{1,2}=2M^2+\frac{1}{2M^2} (2M^2+Q^2)(s-2M^2)
\mp \frac{1}{2M^2}
\sqrt{Q^2(Q^2+4M^2)s(s-4M^2)}\;,
$$
$M$ is the mass of the $u$- and $\bar d-$quarks.
The functions $s_{1,2}(s,Q^2)$ give the kinematically available region
in the plane $(s,s')$.
$G^{u,\bar d}_{E,M}=G^{u,\bar d}_{E,M}(Q^2)$ are Sachs form factors of
$u$- and $\bar d-$quarks.

The non-relativistic limit of
equations (\ref{GqGRIP}) gives the following forms of the
form factors:
$$
G_C(Q^2) = \left(G^u_E(Q^2)+G^{\bar d} _E(Q^2)\right)\,I(Q^2)\;,
\quad G_Q(Q^2) = 0\;,
$$
\begin{equation}
G_M(Q^2) = \frac{M_\rho}{M}\left(G^u_M(Q^2)+G^{\bar d} _M(Q^2)\right)\,
I(Q^2)\;,
\label{GNR}
\end{equation}
$$
I(Q^2) = \int\,k^2\,dk\,k'\,^2\,dk'\,u(k)\,g_{NR}(k,Q^2,k')u(k')\;,
$$
$$
g_{NR}(k,Q^2,k') = \frac{1}{k\,k'\,Q}\,
\left[\theta\left(k' - \left|k - \frac{Q}{2}\right|\right) -
\theta\left(k' - k - \frac{Q}{2}\right)\right]\;.
$$
Let us note that the obtained result
(\ref{GNR}) coincides with the one derived in standard non-relativistic
calculations for composite system form factors in the impulse approximation
\cite{BrJ76}. So, Eq.~(\ref{GqGRIP}) can be considered as a
relativistic generalization of the equations of
Ref.~\cite{BrJ76}.

\section{$\rho$-meson electromagnetic structure}

Electroweak properties of composite hadron systems were
described in the RHD approach in a number of papers.
The most popular approach in the frame of RHD is the
light--front dynamics
\cite{ChC88,Kei94,CaG95pl,GrK84,Ter76,Sch94,ChC88pl}.
Recently some calculations in the frame of instant form
\cite{KrT93,ChY99}  and point form dynamics
\cite{AlK98} have appeared. The
$\rho$ -- meson electromagnetic structure was calculated in
Refs.~\cite{Kei94,CaG95pl,Kar96,MeS02,Sim02} in the light--front dynamics
approach.

In this section we make use of the results
of the previous sections to
calculate the
$\rho$ -- meson electromagnetic properties.

The $\rho$ -- meson electromagnetic form factors are calculated
using Eqs.~(\ref{GqGRIP}), (\ref{g0C})--(\ref{omega}) in MIA.
The wave functions in sense of RHD in Eq. (\ref{GqGRIP}) at
$J=S=1\;,l=0$ are defined by the following expression (see e.g.
\cite{KrT02}):
\begin{equation}
\varphi(k(s)) =\sqrt[4]{s}\,u(k)\,k\;,\quad
k = \frac{1}{2}\sqrt{s - 4\,M^2}\;
\label{phi(s)}
\end{equation}
and is normalized by the condition:
\begin{equation}
\int\,u^2(k)\,k^2\,dk = 1\;,
\label{norm}
\end{equation}
here $u(k)$ is a model wave function.

For the description of the relative motion of quarks (as in
Ref.~\cite{KrT02} in the case of pion) in
Eq.~(\ref{phi(s)}) the following phenomenological wave functions
are used:

1. A Gaussian or harmonic oscillator (HO) wave function
(see, e.g., Ref.~\cite{ChC88pl})
\begin{equation}
u(k)= N_{HO}\,
\hbox{exp}\left(-{k^2}/{2\,b^2}\right).
\label{HO-wf}
\end{equation}

2. A power-law (PL)  wave function \cite{Sch94}:
\begin{equation}
u(k) =N_{PL}\,{(k^2/b^2 +
1)^{-n}}\;,\quad n = 2\;,3\;.
\label{PL-wf}
\end{equation}

3. The wave function with linear confinement from Ref.~\cite{Tez91}:
\begin{equation}
u(r) = N_T \,\exp(-\alpha r^{3/2} - \beta r)\;,\quad \alpha =
\frac{2}{3}\sqrt{2\,M_r\,a}\;,\quad \beta = M_r\,b\;,
\label{Tez91-wf}
\end{equation}
here $a\;,b$ are the parameters of linear and Coulomb parts of
potential respectively, $M_r$ is the reduced mass of the
two--particle system.

For Sachs form factors of quarks
in Eqs. (\ref{g0Q}) we have:
\begin{equation}
G^{q}_{E}(Q^2) =
e_q\,f_q(Q^2)\;,\quad G^{q}_{M}(Q^2) = (e_q + \kappa_q)\,f_q(Q^2)\;,
\label{q ff}
\end{equation}
where $e_q$ is the quark charge and $\kappa_q$ is
the quark anomalous magnetic moment.  For $f_q(Q^2)$
the form of Ref.~\cite{KrT98} is used:
\begin{equation}
f_q(Q^2) =
\frac{1}{1 + \ln(1+ \langle r^2_q\rangle Q^2/6)}\;.
\label{f_qour}
\end{equation}
Here $\langle r^2_q\rangle$ is the mean square radius (MSR) of constituent
quark.

The details describing the cause of the choice (\ref{f_qour})
for the function $f_q(Q^2)$
can be found in Ref.~\cite{KrT98} (see also Ref.~\cite{KrT02})
and are based on the fact that this form gives the asymptotics
of the pion form factor as $Q^2\to\infty$ that coincides with
the QCD asymptotics \cite{MaM73}.

So, for the calculations we use a standard set of parameters of
constituent quark model.
The structure of the constituent quark is described by the following
parameters:
$M_u = M_{\bar d} = M$ is the constituent quark mass,
$\kappa_u\;,\;\kappa_{\bar d}$ are the constituent quarks
anomalous magnetic moments,
$\langle\,r^2_u\rangle = \langle\,r^2_{\bar d}\rangle = \langle\,r^2_q\rangle$
is the quark MSR.
The interaction of quarks in
$\rho$ meson is characterized by wave functions
(\ref{HO-wf}) -- (\ref{Tez91-wf}) with the parameters
$b\;,\;\alpha\;,\;\beta$.

The parameters were fixed in our calculation as follows. We use
$M$=0.25 GeV \cite{KrT01} for the quark mass.
The quark anomalous magnetic moments enter the
equations through the sum ($\kappa_u + \kappa_{\bar d}$) and we take
$\kappa_u + \kappa_{\bar d}$ = 0.09 in natural units following
\cite{Ger95}.

For the quark MSR we use the relation
$\langle r^2_q\rangle \simeq 0.3 /M^2$
\cite{CaG96,VoL90}.

The parameter $b$ in the Coulomb part of the
potential in Eq.~(\ref{Tez91-wf}) is
$b$ = 0.7867.
This gives the value $\alpha_S$ = 0.59 for systems of light quarks.
We choose the parameters $b$ in Eqs. (\ref{HO-wf}) and
(\ref{PL-wf}) and $a$ in Eq. (\ref{Tez91-wf}) so as to fit the
MSR of $\rho$ meson.

The $\rho$ -- meson MSR is given by
the equation $\langle r^2_\rho\rangle - \langle r^2_\pi\rangle$
= 0.11$\pm$0.06 fm$^2$ \cite{CaG96,VoL90}.  For the pion MSR the
experimental data \cite{Ame84} is taken:
$\langle r^2_\pi\rangle^{1/2}$ = 0.657$\pm$0.012 fm.

We use the following relation for the $\rho$-meson MSR:
\begin{equation}
\langle\,r^2_\rho\rangle = -6\,G'_{C}(0)\;.
\label{MSR}
\end{equation}

The $\rho$ -- meson mass in Eq.~(\ref{GqGRIP}) is taken from
Ref.~\cite{Abe99}:  $M_c = M_\rho$ = 763.0$\pm$1.3 MeV.

The magnetic $\mu_\rho$ and the quadrupole $Q_\rho$
moments of $\rho$ meson were calculated using the relations
\cite{ArC80}:
\begin{equation}
G_M(0) = \frac{M_\rho}{M}\,\mu_\rho\;,\quad
G_Q(0) = M_\rho^2\,Q_\rho\;.
\label{stat}
\end{equation}
The static limit in Eqs.~(\ref{GqGRIP}) gives the following
relativistic expressions for moments:
\begin{equation}
\mu_\rho = \frac{1}{2}\,\int_{2M}^\infty\,d\sqrt{s}\,\frac{\varphi^2(s)}
{\sqrt{s - 4\,M^2}}\,\left\{1 - L(s) + (\kappa_u + \kappa_{\bar d})
\left[1 - \frac{1}{2}\,L(s)\right]\right\}\;,
\label{mu}
\end{equation}
\begin{equation}
Q_\rho = -\,\int_{2M}^\infty\,d\sqrt{s}\,\frac{\varphi^2(s)}
{\sqrt{s}}\,
\left[\frac{M}{\sqrt{s} + 2\,M} + \kappa_u + \kappa_{\bar d}\right]
\,\frac{L(s)}{4M\sqrt{s - 4M^2}}\;,
\label{Q}
\end{equation}
$$
L(s) = \frac{2\,M^2}{\sqrt{s - 4\,M^2}\,(\sqrt{s} + 2\,M)}\,\left[
\frac{1}{2\,M^2}\sqrt{s\,(s - 4\,M^2)} + \right.
$$
$$
+ \left.
\ln\,
\frac{\sqrt{s} - \sqrt{s - 4\,M^2}}{\sqrt{s} + \sqrt{s - 4\,M^2}}\right]\;.
$$
Let us note that the nonzero
$\rho$ -- meson quadrupole moment appears due to the
relativistic effect of Wigner spin rotation of quarks only.
So, measuring of the quadrupole moment
can be a test of the relativistic invariance in the confinement
region.

The Wigner rotation contributes to the magnetic moment, too.
Without spin rotation
($\omega_1 = \omega_2$ = 0 in (\ref{omega})) we have for the
magnetic moment:
\begin{equation}
\tilde\mu_\rho = \frac{1}{2}(1 + \kappa_u + \kappa_{\bar d})\,
\int_{2M}^\infty\,d\sqrt{s}\,
\frac{\varphi^2(s)}
{\sqrt{s - 4\,M^2}}\,\left\{1 - \frac{1}{2}\,L(s)\right\}\;.
\label{muwsr}
\end{equation}

To estimate the contribution of Wigner rotation we have calculated
the $\rho$-meson MSR $\langle\,\tilde r_\rho^2\rangle$ without spin rotation
($\omega_1 = \omega_2$ = 0 in Eqs.  (\ref{GqGRIP}), (\ref{g0C}),
(\ref{omega}), (\ref{MSR})).

To estimate the contribution of the relativistic effects to the
$\rho$ -- meson electromagnetic structure the
non-relativistic calculation of electromagnetic form factors, MSR
$\langle\,r^2_{NR}\rangle$
and magnetic moment $\mu_{\rho\,NR}$ was performed using Eqs.
(\ref{GNR}), (\ref{MSR}), (\ref{stat}).

The calculation of the non-relativistic magnetic moments using
Eq.~(\ref{stat}) gives the following result which does not
depend on the model wave functions:
\begin{equation}
\mu_{\rho\,NR} = 1 + \kappa_u + \kappa_{\bar d}\;.
\label{muNR}
\end{equation}
The values of parameters being fixed before we
obtain $\mu_{\rho\,NR}$ = 1.09.

The results of the calculation of the
$\rho$ -- meson statical moments are given in the Table I.
The relativistic results for the same parameters but without Wigner spin
rotation ($\omega_1 = \omega_2$ = 0 in Eq. (\ref{omega})),
as well as of non-relativistic calculation are given, too.
The number of significant digits is
chosen so as to demonstrate the extent of the model
dependence of the calculations.

\begin{table}
\caption{The $\rho$ -- meson statical moments obtained
with the different model wave functions
\protect(\ref{HO-wf}) -- \protect(\ref{Tez91-wf}).
$\langle\,r^2_{NR}\rangle$ is
nonrelativistic MSR, $\langle\,\tilde
r_\rho^2\rangle$ is relativistic MSR without spin rotation
contribution, $\mu_\rho$ is relativistic magnetic moment
(\ref{mu}), $\tilde\mu_\rho$ is relativistic magnetic moment
without spin rotation (\ref{muwsr}),
$Q_\rho$ is quadrupole moment (\ref{Q}).
The wave functions parameters are obtained from the fitting of
$\rho$ -- meson MSR obtained through the relativistic
calculation with spin rotation,
$\langle r^2_\rho\rangle - \langle
r^2_\pi\rangle$ = 0.11$\pm$0.06 fm$^2$ \protect\cite{CaG96,VoL90}. The pion
MSR was
taken from the experimental data
\protect\cite{Ame84}:  $\langle r^2_\pi\rangle^{1/2}$ = 0.657$\pm$0.012 fm.
The magnetic moments are in natural units, the quadrupole
moments and MSR are in fm$^2$.
The parameters $b$ in \protect(\ref{HO-wf}) and \protect(\ref{PL-wf}) --
are in GeV and $a$ in \protect(\ref{Tez91-wf}) -- in GeV$^2$.
The values of other parameters are given in the text.
The number of the significant digits is chosen
so as to demonstrate the extent of model dependence
of calculations.}
\label{tab:1}
\begin{tabular}{ccccccc}
\hline\hline\noalign{\smallskip}
~~Wave~~       &       &       &       &       &       &       \\
~~functions~~  &$b\;,a$&$\langle\,r^2_{NR}\rangle$
               &$\langle\,\tilde r_\rho^2\rangle$
               &$\mu_\rho$&$\tilde \mu_\rho$&$Q_\rho$\\
\noalign{\smallskip}\hline\noalign{\smallskip}
(\ref{HO-wf})    & 0.231 & 0.275 & 0.731 & 0.852 & 0.966 & -0.0065\\
%\hline
(\ref{PL-wf}) n=2& 0.302 & 0.319 & 0.711 & 0.864 & 0.972 & -0.0059\\
%\hline
(\ref{PL-wf}) n=3& 0.430 & 0.305 & 0.710 & 0.866 & 0.973 & -0.0061\\
%\hline
(\ref{Tez91-wf}) & 0.028 & 0.301 & 0.711 & 0.865 & 0.973 & -0.0061\\
\noalign{\smallskip}\hline\hline
\end{tabular}
\end{table}

As one can see from the table the contributions of the spin
rotation to the magnetic moment and to MSR depend weakly on the
model for the quarks interaction in
$\rho$ meson. This contribution to the MSR is
24\%--26\% and is negative. This result differs from that of the
paper \cite{Kei94} where the contribution of spin rotation
(Melosh rotation) to MSR calculated in the frame of
light--front dynamics is positive.

The spin rotation
contribution to the magnetic moment in our calculations is
11\%--12\% and is negative, too.
The total relativistic
corrections to MSR in our approach are positive and enlarge the
non-relativistic value essentially -- almost twice in the
case of the model
(\ref{HO-wf}) and for 70\% -- 80\% for the models (\ref{PL-wf}),
(\ref{Tez91-wf}).

The total relativistic corrections for the
magnetic moment as compared to the non-relativistic result
(see Eq.~(\ref{muNR})) are negative and have the value of
21\%--22\%.  Let us note, that in the light--front dynamics
approach \cite{CaG95pl} a different result was obtained: the
positive relativistic correction of the value of 10\% to the
magnetic moment.

The results of calculations for the $\rho$ -- meson electromagnetic form
factors are represented in Fig.1--4.

\begin{figure}[htbp]%\vspace*{-3.0cm}
\epsfxsize=0.9\textwidth
\centerline{\psfig{figure=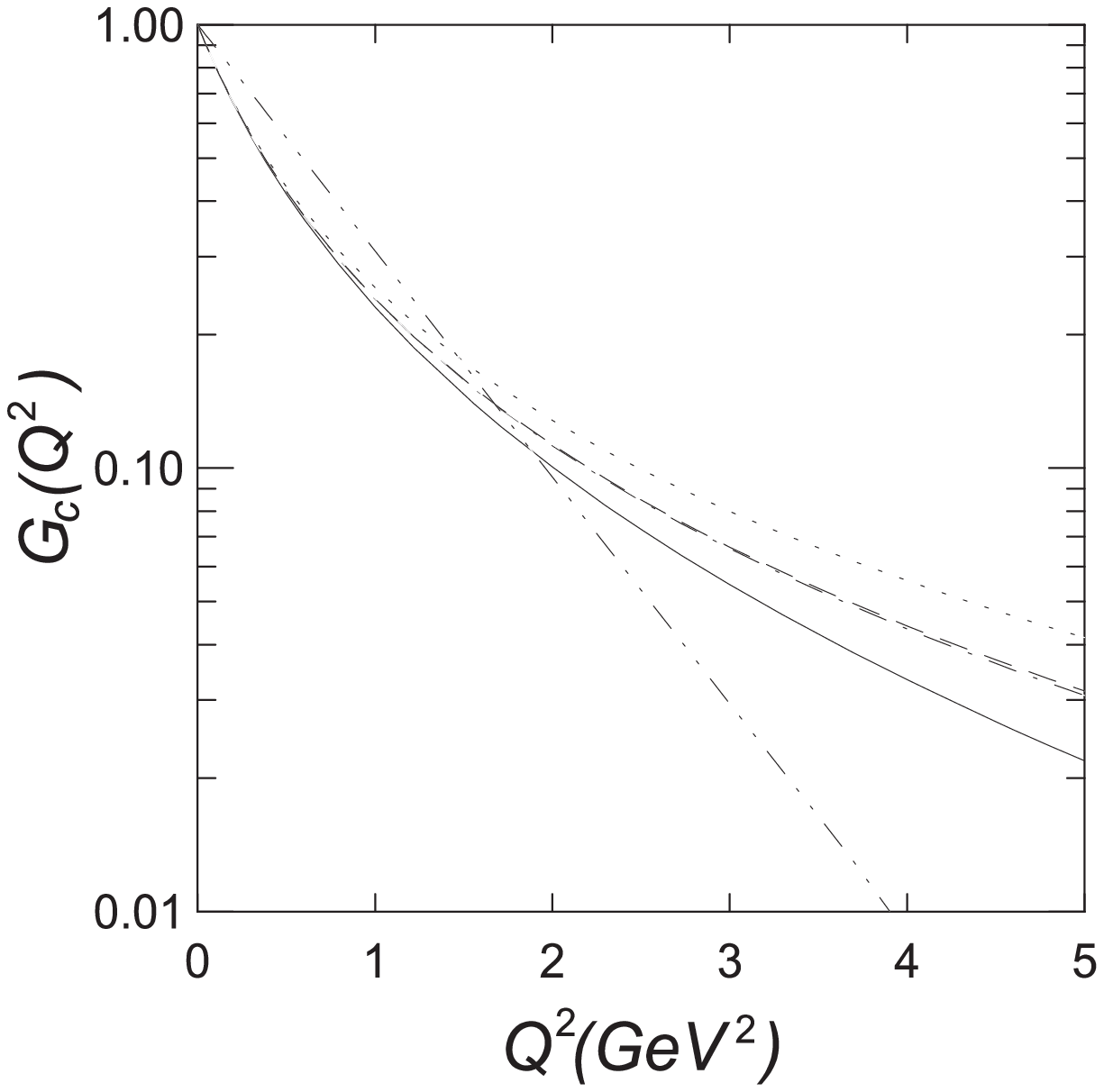,height=8.6cm,width=8.6cm}}
\vspace{0.3cm}
\caption{The results of the calculations of the
$\rho$-meson charge form factor with different model wave
functions. The values of the parameters are given in the
Table I. The solid line represents the relativistic calculation with the wave
function (\ref{HO-wf}), the dashed line -- with (\ref{PL-wf})
for $n =$ 3, dash--dot--line -- with (\ref{Tez91-wf}), dotted
line -- with (\ref{PL-wf}) for $n =$ 2, dot--dot--dash-line --
the non-relativistic calculation with (\ref{HO-wf}).}
\label{fig:1}
\end{figure}

\begin{figure}[htbp]%\vspace*{-3.0cm}
\epsfxsize=0.9\textwidth
\centerline{\psfig{figure=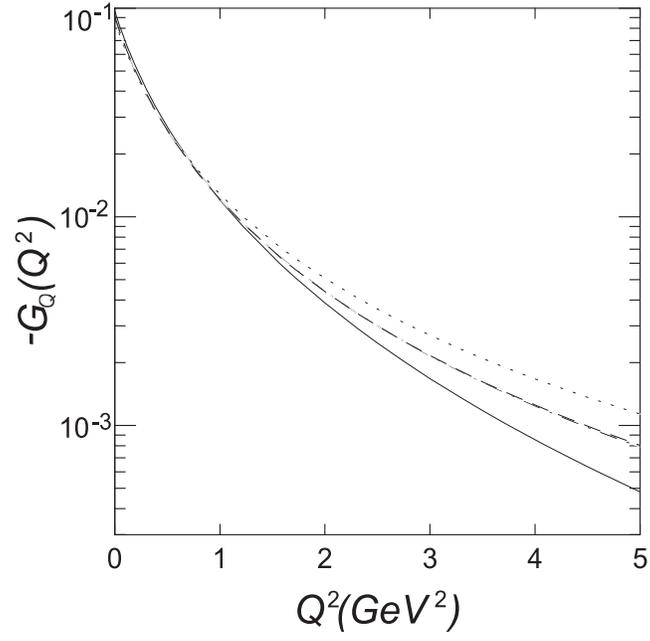,height=8.6cm,width=8.6cm}}
\vspace{0.3cm}
\caption{The results of the calculations of the
$\rho$-meson quadrupole form factor with different model wave
functions, legend as in Fig.1}
\label{fig:2}
\end{figure}

\begin{figure}[htbp]%\vspace*{-0.8cm}
\epsfxsize=0.9\textwidth
\centerline{\psfig{figure=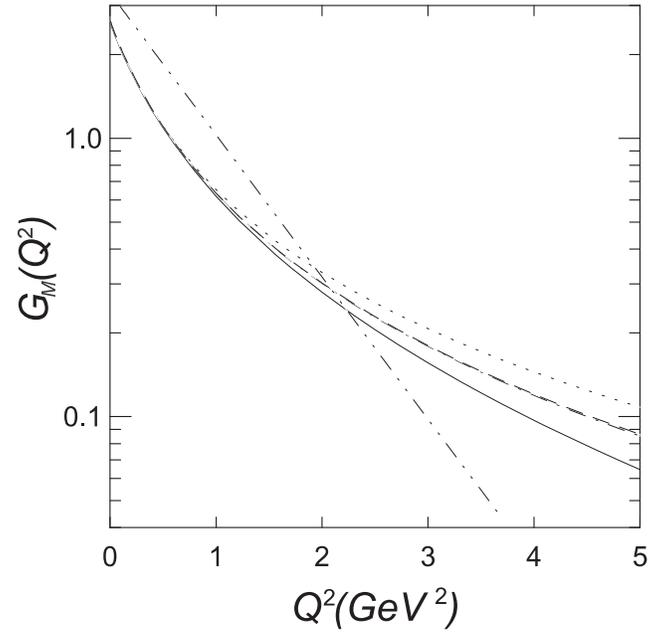,height=8.6cm,width=8.6cm}}
\vspace{0.3cm}
\caption{The results of the calculations of the
$\rho$-meson magnetic form factor with different model wave
functions, legend as in Fig.1}
\label{fig:3}
\end{figure}

\begin{figure}[htbp]%\vspace*{-3.0cm}
\epsfxsize=0.9\textwidth
\centerline{\psfig{figure=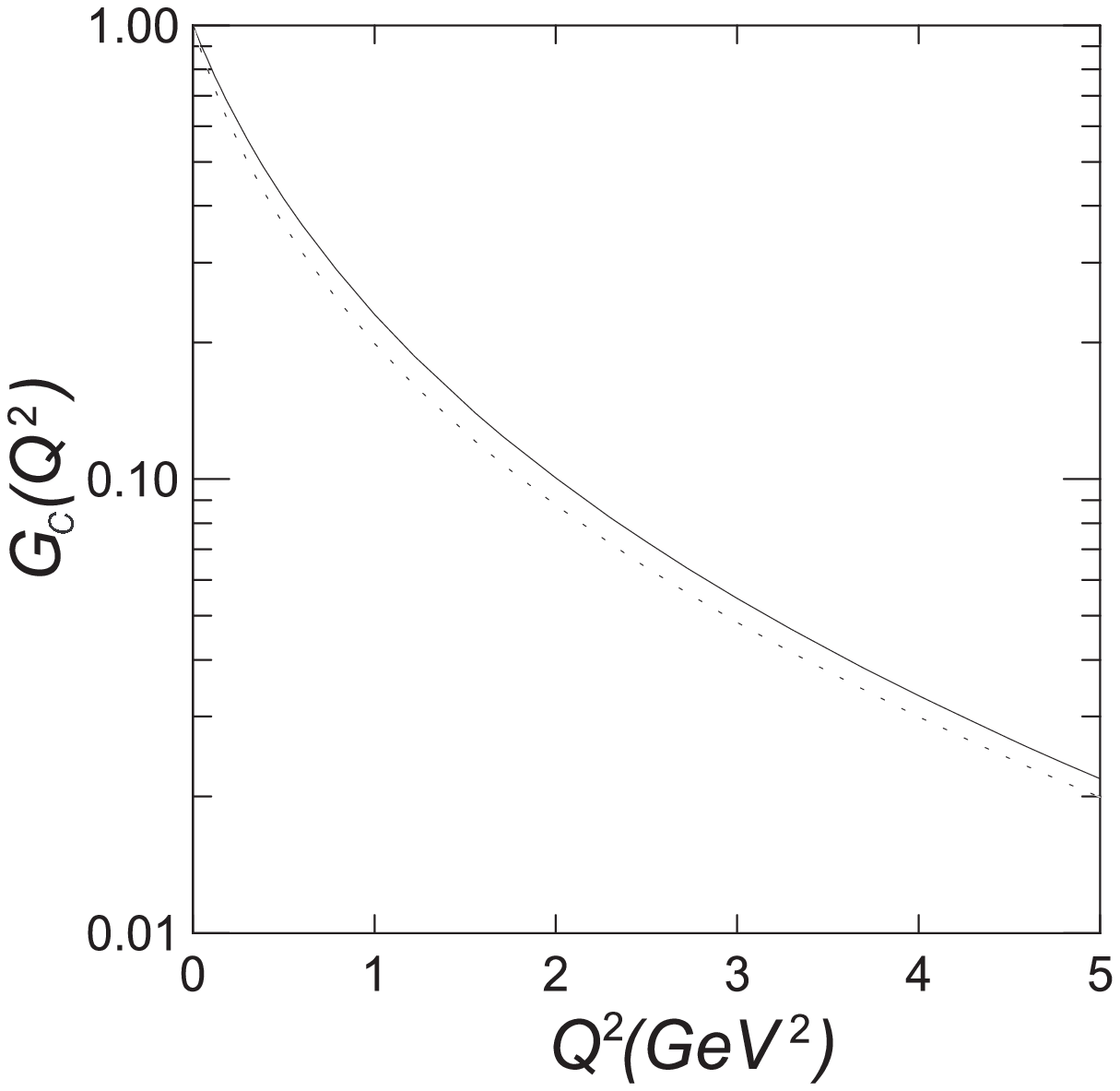,height=8.6cm,width=8.6cm}}
\vspace{0.3cm}
\caption{The contribution of the relativistic spin rotation
effect. The results of the relativistic calculation of the
$\rho$-meson charge form factor with the wave function
(\ref{HO-wf})  using the same parameters as in Fig.1.
Solid line represents the relativistic calculation with spin
rotation, dotted line -- the relativistic calculation without spin
rotation.}
\label{fig:4}
\end{figure}

Let us note that our charge form factor has no dip in contrast with
the results of the paper \cite{CaG95pl}.
The relativistic corrections in our approach diminish
essentially the rate of the decreasing of the charge and
magnetic form factors at large values of momentum transfer. We
demonstrate in the figures the case of the model (\ref{HO-wf})
with the exponential decreasing of the nonrelativistic form
factors with the increasing $Q^2$. The nonrelativistic
quadrupole form factor is zero in the absence of the $D$--state
in the two--particle system.

In Fig.4 the contribution of Wigner rotation of quark spins
to the $\rho$ -- meson charge form factor is shown. This
contribution depends weakly on the momentum transfer in the
range from 1 to  5 GeV$^2$ and its value is approximately 10\%.

\section{Conclusion}

The electromagnetic form factors, quadrupole and magnetic moments, and
MSR of the $\rho$-meson were calculated in the framework of the instant
form of relativistic Hamiltonian dynamics (RHD).

The special method of construction of the electromagnetic current matrix
elements for the relativistic two--particle composite systems
with nonzero total angular momentum is used to obtain the integral
representation for the electromagnetic form factors.

The modified impulse approximation (MIA) --- with the physical content
of the relativistic impulse approximation --- is formulated in
terms of reduced matrix elements on Poincar\'e group. MIA conserves
Lorentz covariance of electromagnetic current and the
current conservation law.

A reasonable
description of the static moments and the electromagnetic form
factors of $\rho$ meson is obtained in the developed formalism.
A number of relativistic effects are
obtained, for example, the nonzero quadrupole moment (in the
case of $S$ state) due to the relativistic Wigner spin rotation.

So, it is shown that the instant form
of RHD can be used to obtain an adequate description of
the electroweak properties of composite systems with nonzero
total angular momentum.

This work was supported in part by the Program "Russian
Universities--Basic Researches" (Grant No. 02.01.013) and Ministry of
Education of Russia (Grant No. E02-3.1-34).

\end{document}